\documentclass[fleqn,10pt]{wlscirep}

\usepackage{graphicx}
\usepackage{color}
\usepackage{multirow}
\usepackage{amsmath,amssymb}

\title{Thermal vacancy formation energies of random solid solutions}

\author[1,*]{H. B. Luo}
\author[2]{Q. M. Hu}
\author[1]{J. Du}
\author[1]{A. R. Yan}
\author[3]{J. P. Liu}
\affil[1]{Key Laboratory of Magnetic Materials and Devices, Ningbo Institute of Materials Technology and Engineering, Chinese Academy of Sciences, Ningbo 315201, China}
\affil[2]{Shenyang National Laboratory for Materials Science, Institute of Metal Research, Chinese Academy of Sciences, 72 Wenhua Road, Shenyang 110016, China}
\affil[3]{Department of Physics, University of Texas at Arlington, Arlington, TX 76091, USA}

\affil[*]{hubin.luo@gmail.com}

\begin{abstract}
Vacancy mechanism plays a dominant role in the atomic migration when a close-packed disordered alloy undergoes ordering transition. However, the calculation of thermal vacancy formation energies (VFEs) of random solid solutions is usually cumbersome due to the difficulty in considering various local atomic environments. Here, we propose a transparent way that combines coherent potential approximation and supercell-local cluster expansion to investigate VFEs of random solid solutions. This method is used to study the effects of temperature, strain and magnetism on the VFEs of random $A1$-FePt alloy. The results show that the mean VFE increases with increasing temperature, decreases under (001) in-plane tensile and compressive strains, and can be further reduced by the magnetic excitation. These effects are explained by discussing the dependence of VFE on local atomic environments and the overall bond strength within.
\end{abstract}

\begin{document}

\flushbottom
\maketitle

\thispagestyle{empty}

\section*{Introduction}

In materials science, point defects play an important role in both physical and chemical behavior of solids. To date, much attention has been drawn to the manipulation of the point defects in, e.g., ceramics, solid catalysts and metallic materials. For instance, the self-diffusion of the atoms in a close-packed crystal basically relies on the vacancy mechanism. Hence, a key physical quantity is the vacancy formation energy (VFE) which determines the vacancy concentration.

In a monoatomic crystal, the thermal VFE can be conveniently calculated as the energy difference between a crystal with a vacancy, and a perfect crystal containing the same number of atoms by using first-principles methods \cite{Korzhavyi1999}. However, in solid solutions, the calculation of VFE is not that straightforward because it depends on the local atomic environments that are very complicated in random solid solutions. Besides, the computational load is also heavy if a large supercell is constructed to describe a disordered system. Despite these, some efforts have still been made to calculate the VFE (or enthalpy) in random solid solutions.

A convenient and efficient approach to treat the solid solutions is to introduce an effective medium such as that defined by the coherent potential approximation (CPA) \cite{Soven1967,Gyorffy1972}. The VFE can thus be calculated in a way similar to that in a monoatomic crystal \cite{Delczeg2012}. In this regard, the CPA calculation assumes that the vacancies are generated within an average local atomic environment. However, the generation of vacancy should be energetically optional if detailed local atomic environments in a disordered system are considered. A better method is to take advantage of the cluster expansion (CE) to obtain the VFEs in any given local atomic environments based on a certain number of energies from first-principles supercell calculations, such as the work by Zhang and Sluiter \cite{Zhang2015}. However, CE is very complicated beyond binary alloy because there are many coefficients for fitting \cite{Sanchez1984,Wolverton1994}. Actually, the CE approach has to treat at least a ternary case for an alloy with vacancy because the vacancy is considered as an additional species \cite{Lechermann2000,Van2005prl,Zhang2015}. An effective way to simplify the CE approach in this case is to introduce the local cluster expansion (LCE) performed in the vicinity of the vacancies to treat a local configuration-dependent energy, which does not involve additional species, as done by Van der Ven and Ceder \cite{Van2005}. Moreover, since the VFE differs due to different local atomic environments, the way to obtain an appropriate thermal average of VFEs is also very tricky \cite{Zhang2015}.

In this paper, we propose a method that combines the CPA and the supercell-LCE to calculate the VFEs in random solid solutions. This is realized by using the locally self-consistent Green's function (LSGF) method, which maps a supercell onto the effective medium and calculates exactly the charge transfer in the local interaction zone (LIZ) with real distribution of atoms \cite{Abrikosov1997,Ruban2008,Peil2012}. It is also convenient to treat paramagnetism using LSGF with the disordered local moment (DLM) approximation. The proposed method is applied to the FePt alloy, of which the ordered $L1_0$ phase is promising to achieve high-density vertical magnetic recording due to its large uniaxial magnetocrystalline anisotropy \cite{Berry2007,Wiedwald2007,Hsiao2009,Wang2011,Li2011b,Wang2013,Luo2014,Luo2015}. The VFE in this alloy is critical since the vacancy mechanism dominates the transition from the disordered $A1$ phase (without uniaxial magnetocrystalline anisotropy) to the ordered $L1_0$ phase \cite{Wang2013,Chepulskii2005}. The effects of (001) in-plane strains and magnetism on the VFEs are investigated because their importance to the ordering transition was reported experimentally \cite{Hsiao2009,Elkins2005}.

Below, we will first introduce the computational idea of thermally averaged VFE combined with local cluster expansion and the first-principles method we adopted. Then, we present the calculated VFEs of $L1_0$-FePt as well as elemental Fe and Pt, following which, the VFEs of $A1$-FePt under different strains and magnetic states are shown. The effects of strain and magnetism on the VFEs are discussed based on the bond strength within different local environments. Finally, we draw a conclusion.

\section*{Methods}

\subsection*{Calculation of vacancy formation energy}

Generally, the VFE for a certain vacated atom is defined as the energy difference between the crystals with and without a vacancy, using the chemical potential of the vacated atom to compensate for the energy loss,
\begin{equation}{\label{eq_VFE}}
E_{\rm{v}}^{\rm{f}}  =  E^{N-1}-E^N+\mu = \varepsilon + \mu,
\end{equation}
where $N$ is the number of atoms. The generation of thermal vacancies is solely attributed to the thermal effects which does not involve composition change or phase separation, i.e., the vacated atoms are still within the same phase. Hence, $\mu$ should be determined by the energy of the vacancy-free crystal. For a pure metal, it is straightforward that $\mu$ is the energy per atom of the crystal. However, for a disordered alloy, $\mu$ is species-specific and composition-dependent, which is tricky to be calculated appropriately \cite{Belak2015}. Nonetheless, the energy per site of the alloy is known to be a concentration-weighted average of the chemical potentials: $E^N/N=\sum_s c_s\mu_s$ ($s$ indexes the species).

Considering an $A-B$ disordered alloy with $N_A+N_B=N$ atoms, if $n$ vacancies are generated by removing $n_A^{}$ $A$ and $n_B^{}$ $B$ ($n_A^{}/n=c_A^{}$), the total VFE is a sum of all VFEs because vacancies are usually rare in a solid so that interactions between them can be neglected, i.e.,
\begin{equation}{\label{eq_totVFE}}
E_{n\rm{v}}^{\rm{f}}(\textbf{r}_1,\textbf{r}_2,\cdots,\textbf{r}_{n_A},\textbf{r}'_1,\textbf{r}'_2,\cdots,\textbf{r}'_{n_B})=\sum\limits_{i=1}^{n_A} E_{\rm{v}_\emph{A}}^{\rm{f}}(\textbf{r}_i) + \sum\limits_{j=1}^{n_B} E_{\rm{v}_\emph{B}}^{\rm{f}}(\textbf{r}'_j),
\end{equation}
in which $E_{\rm{v}_\emph{A}}^{\rm{f}}(\textbf{r}_i)$ and $E_{\rm{v}_\emph{B}}^{\rm{f}}(\textbf{r}'_j)$ are the VFEs with respect to $A$ and $B$ vacancies at the positions of $\textbf{r}_i$ and $\textbf{r}'_j$, respectively. $\textbf{r}_i$ varies within the positions of $A$ and $\textbf{r}'_j$ varies within those of $B$. Choosing a canonical ensemble for this $n$-vacancy system, the partition function is a sum of the Boltzmann factors over all degrees of freedom of the $n$ vacancies \cite{Greiner1995},
\begin{equation}{\label{eq_npar1}}
Z_n = \sum\limits_{\textbf{r}_i\in\{\textbf{R}_{l}\}}\sum\limits_{\textbf{r}'_j\in\{\textbf{R}'_{k}\}}\exp(-\beta E_{n\rm{v}}^{\rm{f}}),
\end{equation}
in which $\beta=1/(k_{\rm{B}}T)$ ($k_{\rm{B}}$ is the Boltzmann constant and $T$ is the temperature), and $\{\textbf{R}_{l}\}$ and $\{\textbf{R}'_{k}\}$ are the sets of $A$ and $B$ positions, respectively. Using Eqs.~\ref{eq_totVFE} and \ref{eq_npar1}, it is easy to obtain $Z_n=Z_{n_A}Z'_{n_B}$ with $Z_{n_A}$ and $Z'_{n_B}$ being the partition functions for the subsystems of $A$ and $B$ vacancies, respectively. For a noninteracting canonical ensemble, a multi-particle partition function can be further described by using the single-particle partition functions \cite{Greiner1995},
\begin{equation}{\label{eq_npar2}}
Z_n = \frac{1}{n_A!n_B!}{Z_1}^{n_A}{Z'_1}^{n_B}
\end{equation}
with
\begin{equation}{\label{eq_spar1}}
Z_1 = \sum\limits_{i=1}^{N_A}\exp(-\beta E_{{\rm{v}}_{A_i}}^{\rm{f}})
\end{equation}
and
\begin{equation}{\label{eq_spar2}}
Z'_1 = \sum\limits_{j=1}^{N_B}\exp(-\beta E_{{\rm{v}}_{B_j}}^{\rm{f}}).
\end{equation}
The factor $1/n_A!$ ($1/n_B!$) results from the equivalence of different generation sequences of $A$ ($B$) vacancies. The summations in $Z_1$ and $Z'_1$ extend over all $A$ and $B$ sites, respectively.  The mean VFE is then given by
\begin{eqnarray}{\label{eq_mVFEp}}
\overline{E}_{\rm{v}}^{\rm{f}} & = & -\frac{1}{n}\frac{\partial \ln Z_n}{\partial \beta} \nonumber \\
                               & = & -c_A^{}\frac{\partial \ln Z_1}{\partial \beta} - c_B^{}\frac{\partial \ln Z'_1}{\partial \beta} \nonumber \\
                               & = &  c_A^{}\overline{E}_{{\rm{v}}_A}^{\rm{f}} + c_B^{}\overline{E}_{{\rm{v}}_B}^{\rm{f}},
\end{eqnarray}
which shows that the calculation of the mean VFE of an alloy can be reduced to independent calculations of the mean VFEs for $A$ and $B$ vacancies. This equation indicates that it is not necessary to define an effective VFE for an alloy before the statistical calculation of mean VFE.\cite{Van2005,Zhang2015} In an alloy, taking the $A$ sites for instance, generating a vacancy on different sites results in different $E_{{\rm{v}}_A}^{\rm{f}}$. The associated thermal excitation probability is
\begin{equation}{\label{eq_prob}}
P_{A_i} = \frac{\exp(-\beta E_{{\rm{v}}_{A_i}}^{\rm{f}})}{\sum\limits_{i=1}^{N_A}\exp(-\beta E_{{\rm{v}}_{A_i}}^{\rm{f}})}=\frac{\exp(-\beta \varepsilon_{A_i})}{\sum\limits_{i=1}^{N_A}\exp(-\beta \varepsilon_{A_i})}.
\end{equation}
Note that $\mu_A^{}$ is constant so that $e^{-\beta\mu_A}$ can be reduced. It is thus straightforward to have
\begin{equation}{\label{eq_mEA}}
\overline{E}_{{\rm{v}}_A}^{\rm{f}} = \sum\limits_{i=1}^{N_A}E_{{\rm{v}}_{A_i}}^{\rm{f}}P_{A_i}=\overline{\varepsilon}_A + \mu_A^{}.
\end{equation}
Similar relation can also be obtained for $\overline{E}_{{\rm{v}}_B}^{\rm{f}}$. Hence, Eq.~\ref{eq_mVFEp} can be written as
\begin{equation}{\label{eq_mev}}
\overline{E}_{\rm{v}}^{\rm{f}} = c_A^{}\overline{\varepsilon}_A^{} + c_B^{}\overline{\varepsilon}_B^{} + E^N/N.
\end{equation}
To calculate $\overline{\varepsilon}_A^{}$ and $\overline{\varepsilon}_B^{}$, we may transform the sums into integrals by using the $\delta$ function in a way like
\begin{equation}
f(\varepsilon_{i})=\int \delta(\varepsilon - \varepsilon_{i})f(\varepsilon) d\varepsilon,
\end{equation}
which will result in
\begin{equation}{\label{eq_mepA}}
\overline{\varepsilon}_A^{} = \frac{\int g_A(\varepsilon_A)\varepsilon_A\exp(-\beta\varepsilon_A)d\varepsilon_A}{\int g_A(\varepsilon_A)\exp(-\beta\varepsilon_A)d\varepsilon_A}
\end{equation}
and
\begin{equation}{\label{eq_mepB}}
\overline{\varepsilon}_B^{} = \frac{\int g_B(\varepsilon_B)\varepsilon_B\exp(-\beta\varepsilon_B)d\varepsilon_B}{\int g_B(\varepsilon_B)\exp(-\beta\varepsilon_B)d\varepsilon_B}
\end{equation}
with
\begin{equation}{\label{eq_dos}}
  g_{A/B}^{} = \sum\limits_{i=1}^{N_{A/B}} \delta(\varepsilon_{A/B}^{}-\varepsilon_{A_i/B_i}),
\end{equation}
where $g_{A/B}^{}$ is the density of sites for the vacancies with respect to $\varepsilon_{A/B}^{}$ (an analogue to electronic density of states). To obtain the density of sites, the $\delta$ function can be treated with the Gaussian function. Unlike the previous methods to calculate the mean VFE, this approach does not need the chemical potentials or Monte carlo simulation \cite{Zhang2015,Van2005}. The mean VFE is conveniently calculated once we get $g_{A/B}^{}$.

\subsection*{Local cluster expansion}

\begin{figure}
\begin{minipage}[l]{0.9\linewidth}
  \begin{minipage}[l]{0.25\linewidth}
    \includegraphics[width=3.2cm]{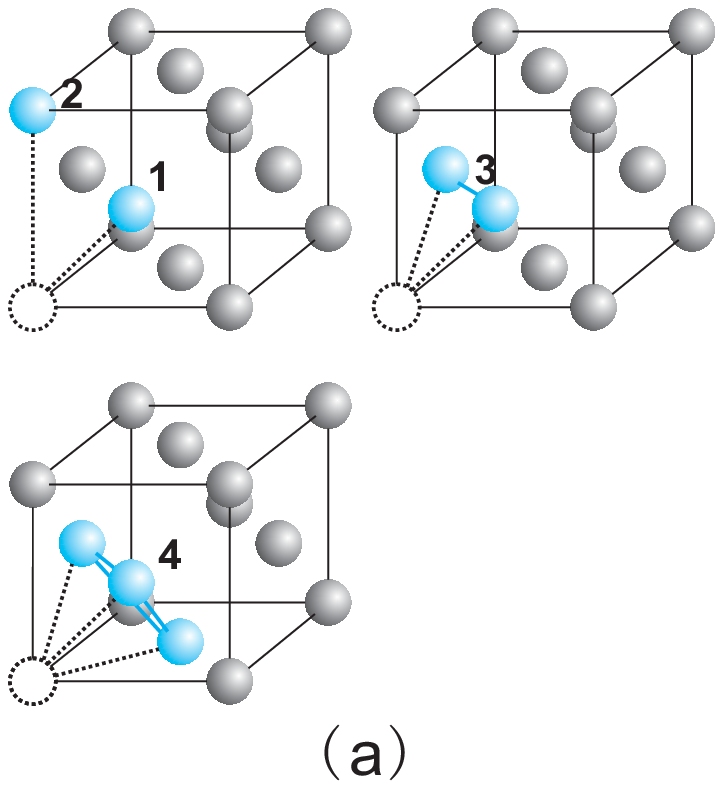}
  \end{minipage}
  \begin{minipage}[r]{0.25\linewidth}
    \includegraphics[width=3.0cm]{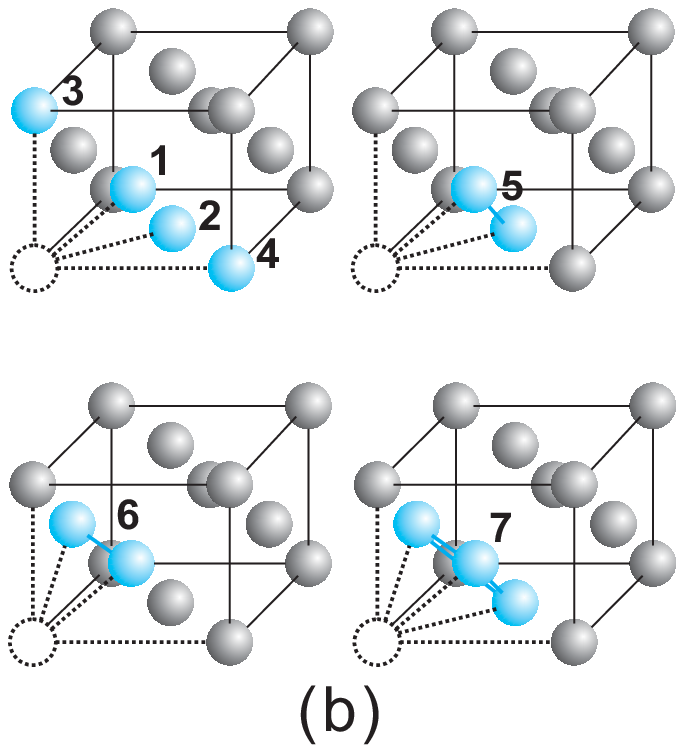}
  \end{minipage}
  \begin{minipage}[l]{0.25\linewidth}
    \includegraphics[width=3.2cm]{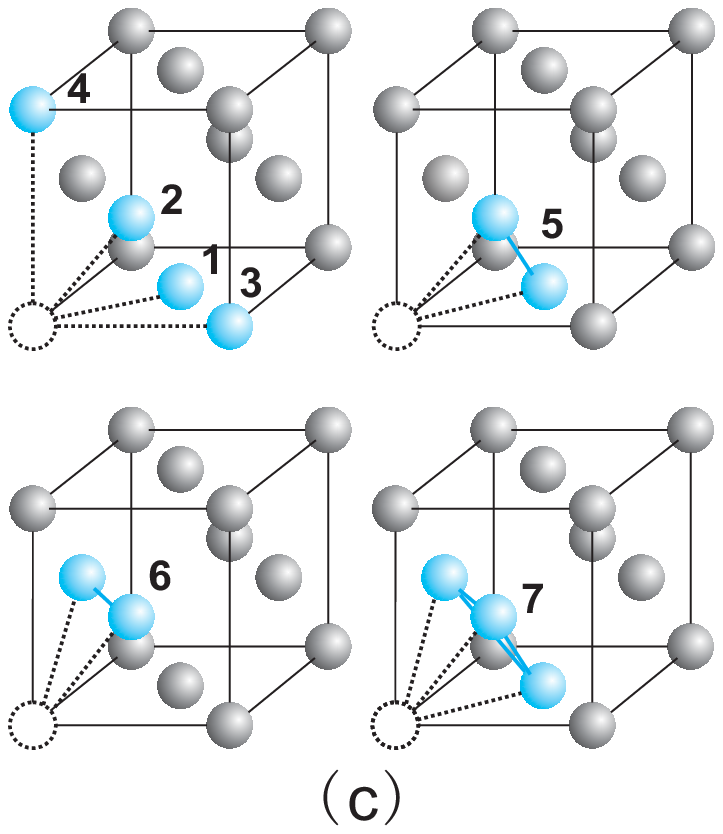}
  \end{minipage}
\end{minipage}
\caption{\label{clst} The local clusters chosen for (a) the unstrained cubic lattice, (b) the lattice with tensile strain ($5\%$) and (c) the lattice with compressive strain ($-5\%$).}
\end{figure}

As its name implies, the LCE is realized by applying cluster expansion around a vacancy to the local configuration-dependent energies \cite{Van2005}. Note that if choosing the energy of free atom as the chemical potential, Eq.~\ref{eq_VFE} is actually to calculate the bond strength between the central atom (to be vacated) and its nearest neighbors (NNs), although it is not reasonable to define the VFE in this way \cite{Zhang2015}. Since the energy of free atom is constant, the variation of $\varepsilon_{A/B}^{}$ depends only on the local atomic environment in the alloy, despite the absolute value of $\varepsilon_{A/B}^{}$ does not make any physical sense. Hence, considering the binary alloy here, $\varepsilon_{\rm{Fe/Pt}}^{}$ can be expanded as
\begin{equation}{\label{eq_lce}}
\varepsilon_{\rm{Fe/Pt}}^{} = J_0 + \sum\limits_{\alpha}J_{\alpha}\Phi_{\alpha}(\vec{\sigma}),
\end{equation}
where $\Phi_{\alpha}(\vec{\sigma})$ is the cluster function corresponding to the specified cluster, $\alpha$, which can be obtained simply by the product of the occupation indexes, i.e., $\Phi_{\alpha}(\vec{\sigma}) = \Pi_{i\in\alpha}\sigma_i$ with $\sigma_i$ taking the value of $1$ ($-1$) when Fe (Pt) occupies the $i$th position in cluster $\alpha$. The coefficients $J_{\alpha}$ are the so-called local effective cluster interactions (LECIs). The local clusters are chosen within the atoms around the vacancy up to the next NNs for the unstrained lattice (cubic) and, due to the lowering of symmetry, up to the fourth NNs for the strained ones, as shown in Fig.~\ref{clst}. This is a good compromise between accuracy and efficiency since the atom-vacancy interactions are very small beyond the next NNs in the face-centered cubic (fcc) metals \cite{Korzhavyi1999}.

To determine the LECIs, we calculate nine $\varepsilon_{\rm{Fe}}^{}$ and nine $\varepsilon_{\rm{Pt}}^{}$ within different local environments in a $4\times4\times4$ (256 atoms) supercell for both ferromagnetic and paramagnetic (DLM) states using first-principles method. The ``disordered" supercell is constructed by optimizing the Warren-Cowley short-range order parameter \cite{Peil2012}. The LECIs are then obtained by least-square fit. Finally, we construct further in this way a $9\times9\times9$ (2916 atoms) supercell and calculate directly the $\varepsilon_{\rm{Fe/Pt}}^{}$ with respect to each site via Eq.~\ref{eq_lce} to obtain $g_{\rm{Fe/Pt}}^{}$. Larger supercells are also tested, but the results have shown to be converged.

\subsection*{Computational details}

In the LSGF calculation, a ``disordered" supercell constructed by optimizing the Warren-Cowley short-range order parameter is mapped onto the CPA medium \cite{Ruban2008,Peil2012}. Around each site, the LIZ contains real distributed atoms and the cluster Dyson equation is calculated exactly within to obtain the charge transfer (outside the LIZ it is the CPA effective medium). Here, the LIZ is chosen to include the next NNs for the unstrained lattice and the fourth NNs for the lattices with strains (tensile or compressive) of $5\%$. This practical method allows us to treat hundreds of atoms without much loss of efficiency.

The LSGF calculation is performed within the exact muffin-tin orbitals (EMTO) formalism based on an improved screened Koringa-Kohn-Rostoker (KKR) method \cite{Vitos2001,Vitos2007}. The scalar-relativistic Green's function is adopted to solve the one-electron Kohn-Sham equation. The one-electron potential is represented by optimized overlapping muffin-tin potential spheres with soft-core approximation. The full-charge density correction is introduced in the total energy calculations \cite{Vitos2007}. The generalized gradient approximation (GGA) parameterized by Perdew et al.\ is employed to describe the electronic exchange-correlation potential \cite{Perdew2008}. The $spdf$ orbitals are included in the EMTO basis sets to construct the wave functions. The Fe-$3d^74s^1$ and Pt-$5d^86s^2$ are treated as valence states. A Gaussian mesh of 16 energy points on a semicircle (1.2 Ry in diameter) comprising the valence states is used for energy integration. The Brillouin Zone of the primitive cell is sampled by special $k$-point technique with a dense mesh of $32\times32\times32$. Local relaxation results in a few percent of difference in VFE but is stable for the same alloy system and thus is not considered here. Besides, the vibrational effect on VFE is also neglected, as is usually done in literature \cite{Glensk2014}.

\section*{Results}

\subsection*{Vacancy formation energy of ordered phase}

We have calculated the VFEs of $L1_0$-FePt as well as the elemental Fe (fcc) and Pt, as shown in Table \ref{vfe_L10}. The EMTO-LSGF calculation obtains a higher $E_{\rm{v}}^{\rm{f}}$ for the ferromagnetic (FM) $L1_0$-FePt than that for the paramagnetic (DLM) one. For the elemental Fe and Pt, EMTO-LSGF yields a large $E_{\rm{v}}^{\rm{f}}$ for Fe (DLM or NM) that is about twice the value of $E_{\rm{v}}^{\rm{f}}$ for Pt. For comparison, except for the DLM cases, the VFE are also calculated by using the pseudopotential plane waves implemented by projector augmented wave (PP/PAW) method performed within the \textsc{VASP} code \cite{Kresse1996}. The results of EMTO-LSGF calculation are larger than those of PP/PAW calculation. This is a general overestimation due to the spherically symmetric potentials adopted by the EMTO method \cite{Korzhavyi1999}. However, a stable acceptable discrepancy does not influence the discussion when we investigate the dependence of VFE on the local atomic environments and external strains. The EMTO-LSGF calculation obtains very close $E_{\rm{v}}^{\rm{f}}$ to that from full-potential calculation for Pt \cite{Korhonen1995}. For Fe and Pt under NM state, the present results of EMTO-LSGF calculation are also consistent with the previous ones of LSGF calculation implemented with the monopole-corrected atomic sphere approximation (ASA) \cite{Korzhavyi1999}.

\begin{table}
\caption{\label{vfe_L10} The VFEs (in eV) of $L1_0$-FePt as well as the elemental Fe and Pt calculated by both EMTO-LSGF and pseudopotential (GGA-PBE) methods. FM, DLM and NM correspond to ferromagnetic, paramagnetic and nonmagnetic states, respectively. Other available $ab~initio$ results are also listed for comparison.}
\begin{minipage}[t]{\textwidth}
\begin{tabular}{lccc} \hline
 \multirow{2}{*}{Material} &  \multicolumn{3}{c}{$E_{\rm{v}}^{\rm{f}}$} \\ \cline{2-4}
                           & EMTO-LSGF & PP/PAW & Refs.   \\ \hline
    FePt (FM)              & 2.299     & 1.823  &         \\
    FePt (DLM)             & 2.231     &        &         \\
    fcc-Fe (DLM)           & 2.694     &        &         \\
    fcc-Fe (NM)            & 3.075     & 2.463  & 2.653\footnote{Ref. \cite{Korzhavyi1999}, ASA+LSGF} \\
    Pt (NM)                & 1.360     & 1.048  & 1.211$^a$, 1.456\footnote{Ref. \cite{Korhonen1995}, FP-LMTO} \\ \hline
\end{tabular}
\end{minipage}
\end{table}

\subsection*{Vacancy formation energy of $A1$-FePt}

\begin{figure*}
\includegraphics[width=5.3cm]{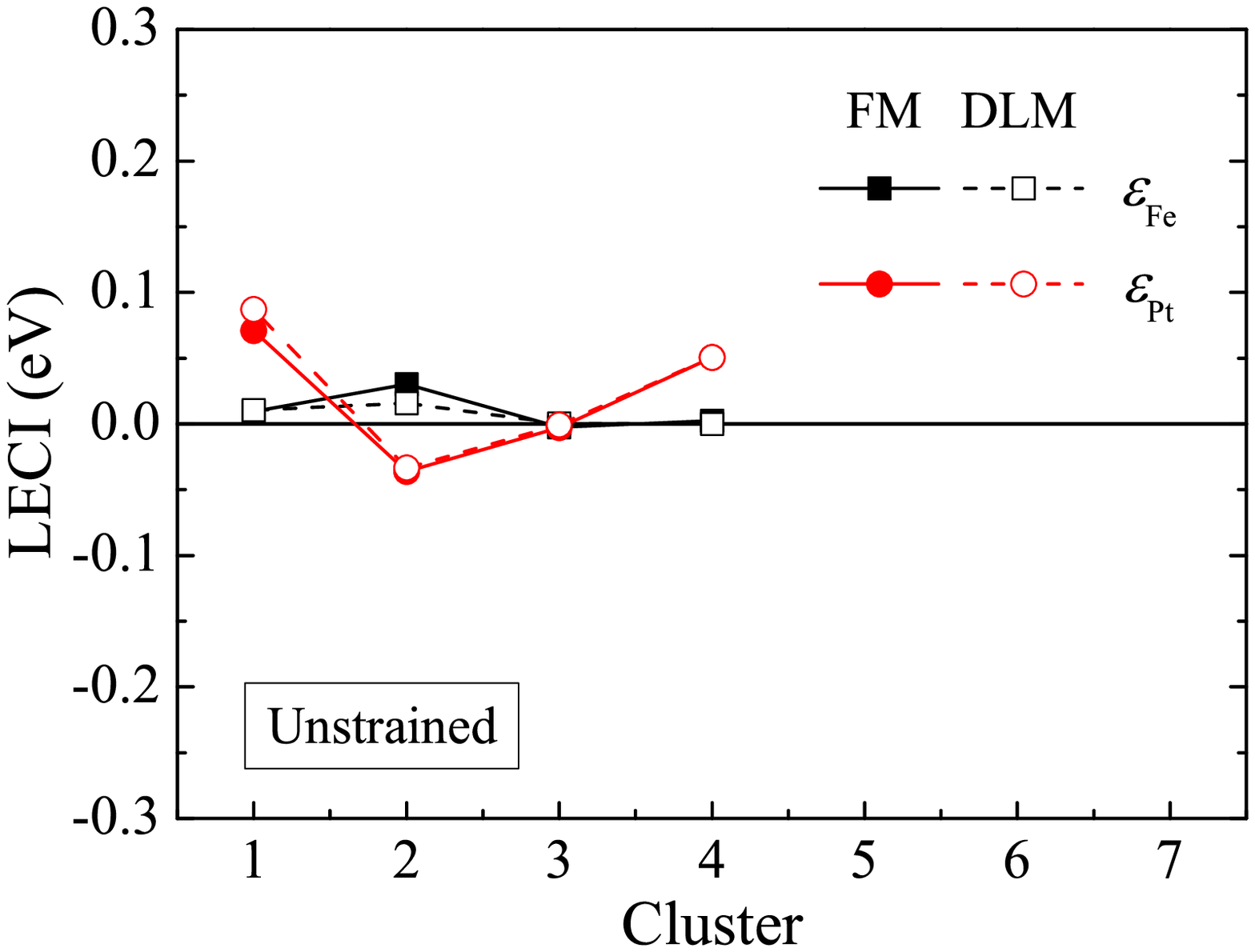}
\includegraphics[width=5.3cm]{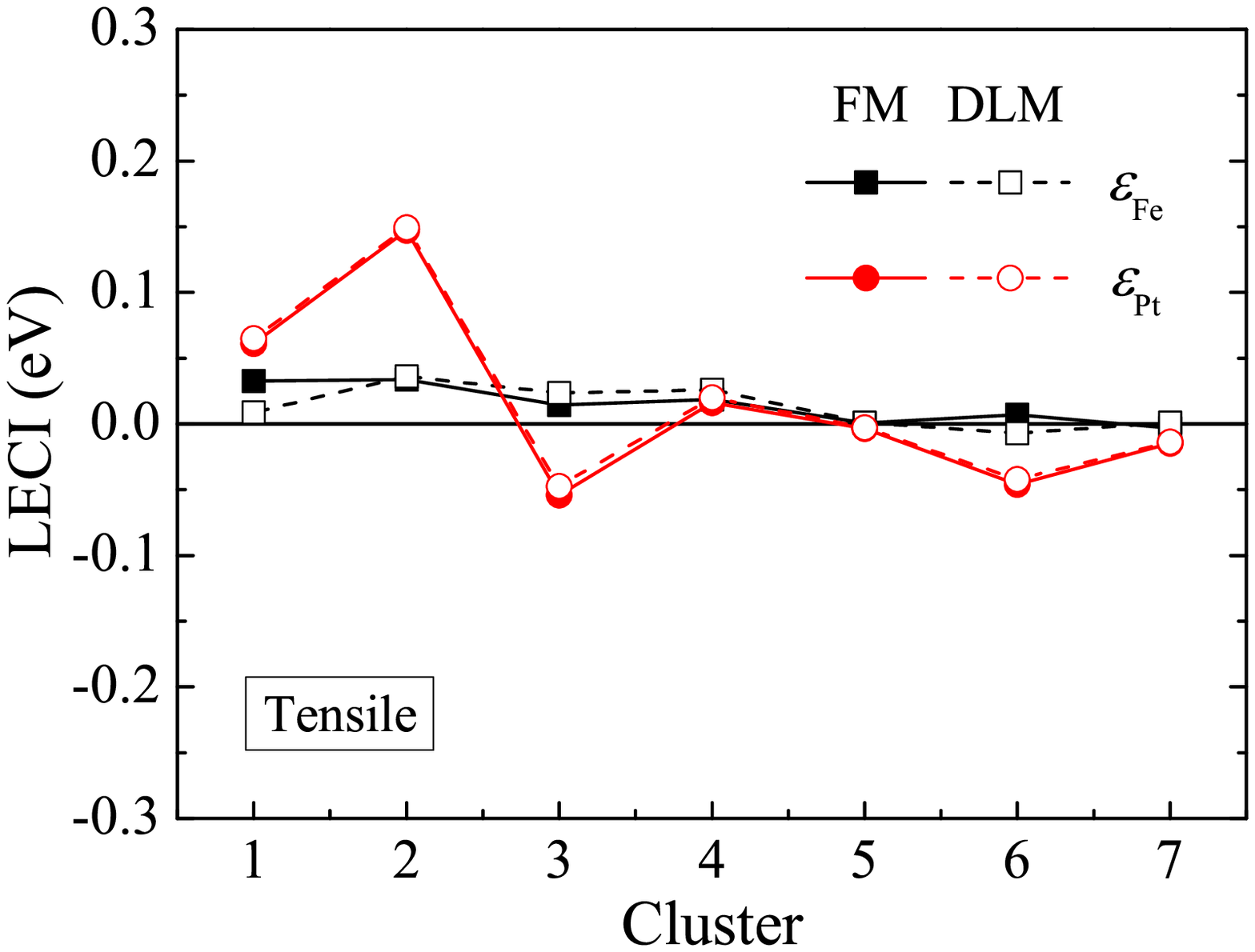}
\includegraphics[width=5.3cm]{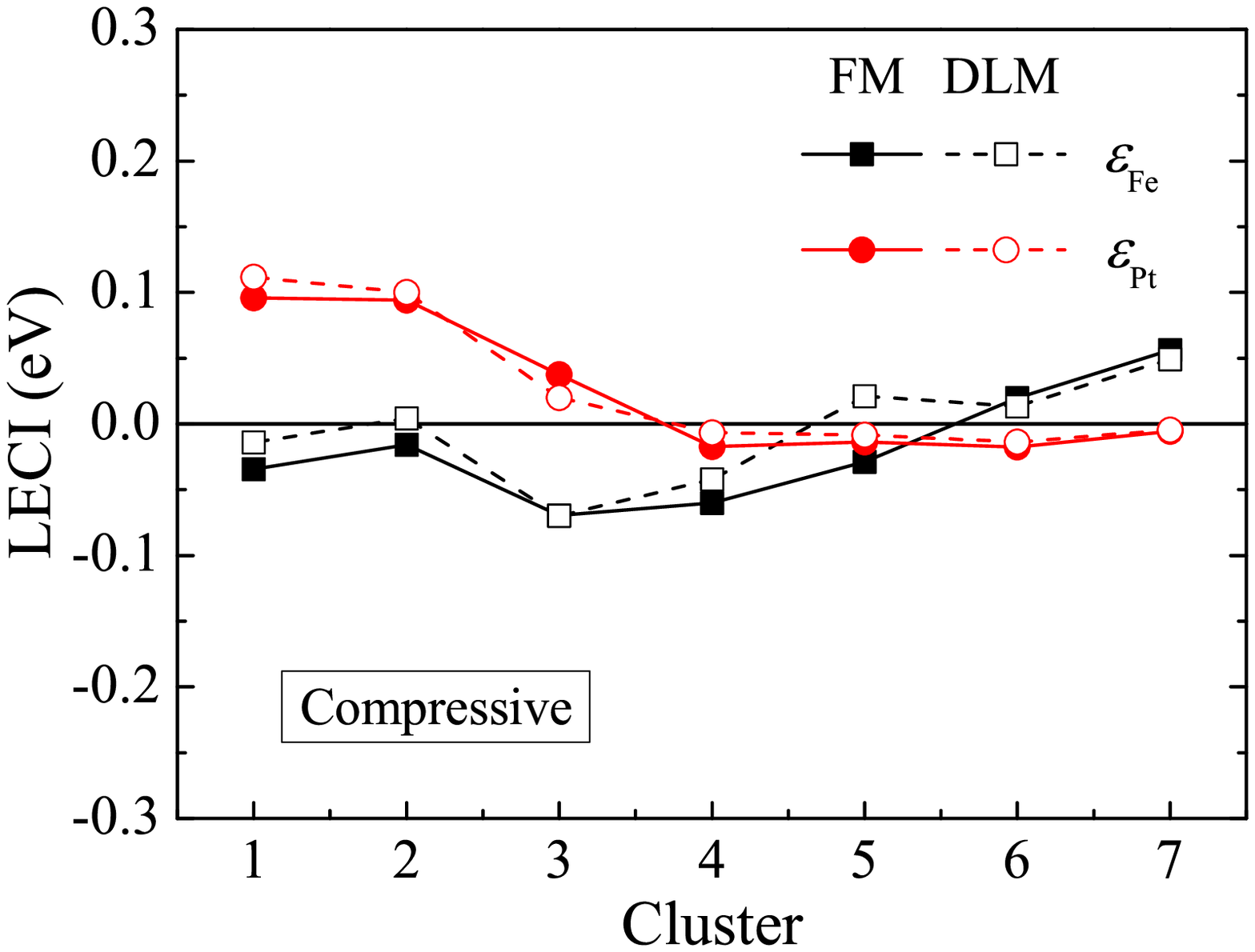}\\
\makebox[2.8cm][l]{}\makebox[5.3cm][l]{\textsf{(a)}} \makebox[5.3cm][l]{\textsf{(b)}} \makebox[3.6cm][l]{\textsf{(c)}}
\caption{\label{leci}  (Color online) The local effective cluster interactions (in eV) for $\varepsilon_{\rm Fe}^{}$ (squares) and $\varepsilon_{\rm Pt}^{}$ (circles) with (a) unstrained lattice, (b) lattice under tensile strain and (c) lattice under compressive strain. The full and open symbols are those corresponding to ferromagnetic (FM) and paramagnetic (DLM) state, respectively.}
\end{figure*}

Since $\varepsilon_{\rm Fe}^{}$ and $\varepsilon_{\rm Pt}^{}$ should be treated separately, we have two sets of LECIs under a given magnetic state (FM or DLM), as shown in Fig.~\ref{leci}. For the same cluster, the LECIs for $\varepsilon_{\rm Pt}^{}$ generally have larger absolute values than those for $\varepsilon_{\rm Fe}^{}$, whereas the LECIs for $\varepsilon_{\rm Fe}^{}$ are more sensitive to the magnetic state. This indicates that $\varepsilon_{\rm Pt}^{}$ has a stronger dependence on the local environment than $\varepsilon_{\rm Fe}^{}$, while $\varepsilon_{\rm Fe}^{}$ is more sensitive to the magnetic state. It can also be found that the LECIs for $\varepsilon_{\rm Fe}^{}$ become more significant when the strains are applied. Hence, a resulting enhanced dependence of $\varepsilon_{\rm Fe}^{}$ on the local environment may also be expected from the strains.

\begin{figure}
\includegraphics[width=5.6cm]{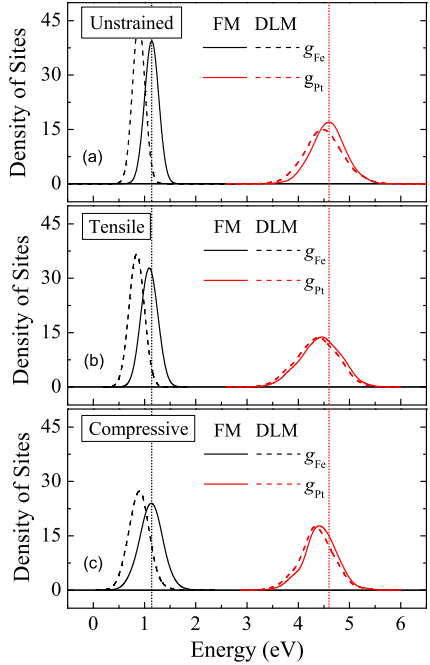}
\caption{\label{dos} (Color online) The normalized density of sites for Fe vacancies ($g_{\rm Fe}^{}$) and Pt vacancies ($g_{\rm Pt}^{}$) distributed on the energies $\varepsilon_{\rm Fe}^{}$ and $\varepsilon_{\rm Pt}^{}$. Two reference energies, 34604 eV and 501825 eV, have been subtracted from $\varepsilon_{\rm Fe}^{}$ and $\varepsilon_{\rm Pt}^{}$, respectively, to make the data be conveniently plotted together for better view. The peaks of the density of sites for unstrained lattice under FM state are marked by the dotted lines.}
\end{figure}

The effects of strain and magnetism on $\varepsilon_{\rm Fe}^{}$ and $\varepsilon_{\rm Pt}^{}$ can be directly observed from the density of sites ($g$) calculated by using Eq.~\ref{eq_dos}, as shown in Fig.~\ref{dos}. Due to a stronger dependence on local atomic environments, $g_{\rm Pt}^{}$ has a larger energy span ($\Delta\varepsilon$) than $g_{\rm Fe}^{}$. With the strains applied, the $\varepsilon_{\rm Fe}^{}$ with respect to $g_{\rm Fe}^{\rm max}$ remains almost unchanged but $\Delta\varepsilon_{\rm Fe}^{}$ increases. In contrast, $g_{\rm Pt}^{}$ shifts to lower energy region under the strains without much change in $\Delta\varepsilon_{\rm Pt}^{}$. When the magnetic state is switched from FM to DLM, $\Delta\varepsilon_{\rm Fe}^{}$ narrows and shifts to lower energy region whereas $\Delta\varepsilon_{\rm Pt}^{}$ does not vary much. Further discussion will be given later.

\begin{figure}
\includegraphics[width=6.3cm]{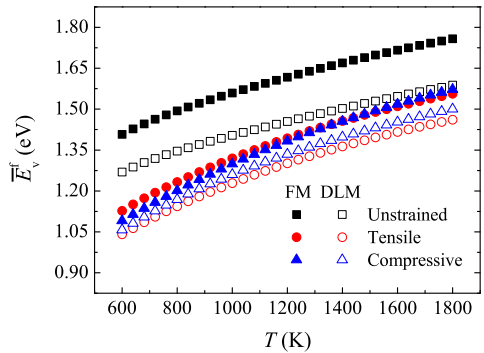}
\caption{\label{evac} (Color online) The temperature dependence of mean VFEs (in eV) under different strains and magnetic states.}
\end{figure}

The mean VFEs ($\overline{E}_{\rm{v}}^{\rm{f}}$) are thus calculated according to Eqs.~\ref{eq_mev}--\ref{eq_dos} by using the obtained $g_{\rm Fe}^{}$ and $g_{\rm Pt}^{}$. The temperature range is chosen to vary from 600 to 1800 K, as shown in Fig.~\ref{evac}, which covers the annealing temperatures that are usually adopted in experiments \cite{Berry2007}. For the alloy with the same strain and magnetic state, $\overline{E}_{\rm{v}}^{\rm{f}}$ increases with temperature. This is because the vacancies with higher $E_{\rm{v}}^{\rm{f}}$ (larger $\varepsilon_{\rm Fe}^{}$ or $\varepsilon_{\rm Pt}^{}$) can be generated when temperature is raised. A similar temperature dependence of vacancy formation enthalpy of Cu-Ni has also been found by Zhang and Sluiter \cite{Zhang2015}. At the same temperature, both the lattice distortion and magnetic excitation lower the $\overline{E}_{\rm{v}}^{\rm{f}}$, and a reduction of $\overline{E}_{\rm{v}}^{\rm{f}}$ up to $20\%$ is found for the alloy with tensile strain under DLM state compared to the unstrained one under FM state. This stems from the downward movement of the lower energy boundary of $g_{\rm Fe}^{}$ and $g_{\rm Pt}^{}$ following the lattice distortions and magnetic excitation (cf. Fig.~\ref{dos}), because it is the vacancies with lower $\varepsilon_{\rm Fe}^{}$ and $\varepsilon_{\rm Pt}^{}$ that are easier to be thermally generated and thus their contributions dominate $\overline{E}_{\rm{v}}^{\rm{f}}$.

For comparison, we have also calculated the $E_{\rm{v}}^{\rm{f}}$ with average local atomic environment using the EMTO-CPA. A result of 2.177 eV for $A1$-FePt (FM) is obtained which is remarkably higher than those presented in Fig.~\ref{evac}. Note that we can also calculate the VFE with an averaged local atomic environment from Fig.~\ref{dos} using the $\varepsilon_{\rm Fe/Pt}^{}$ with respect to $g_{\rm Fe/Pt}^{\rm max}$. The resulting $E_{\rm{v}}^{\rm{f}}$ of 2.163 eV is quite close to the one obtained from CPA calculation, which verifies the accuracy and predictive power of this method. Experimentally, thin FePt films with epitaxial strains in (001) plane generally have lower ordering transition temperature than the bulk ones \cite{Hsiao2009,Elkins2005}, which may be an evidence of our findings that strains may introduce more vacancies due to a lower mean VFE at a given temperature under the same magnetic state.

\section*{Discussion}

\begin{figure}
\includegraphics[width=7.3cm]{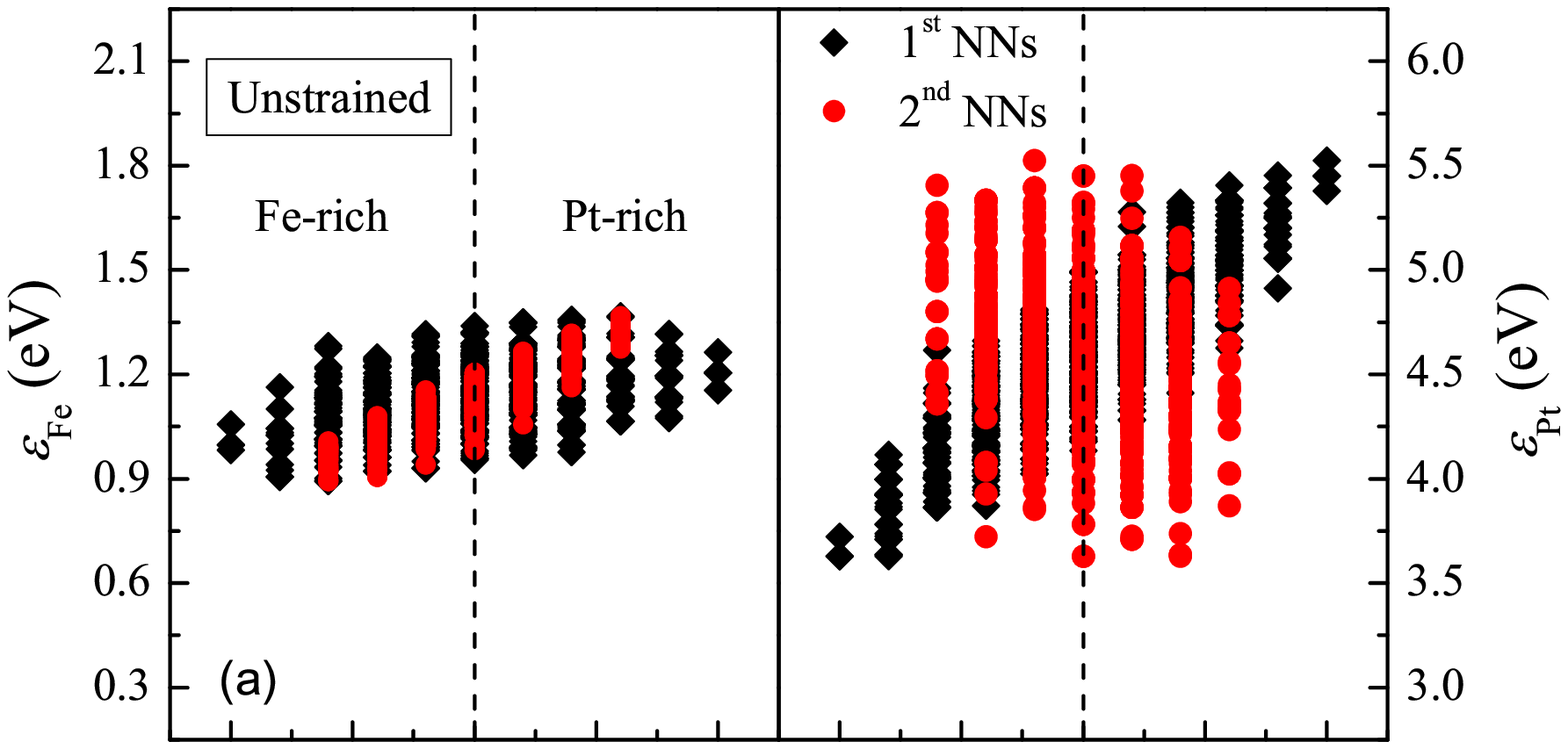}\\
\includegraphics[width=7.3cm]{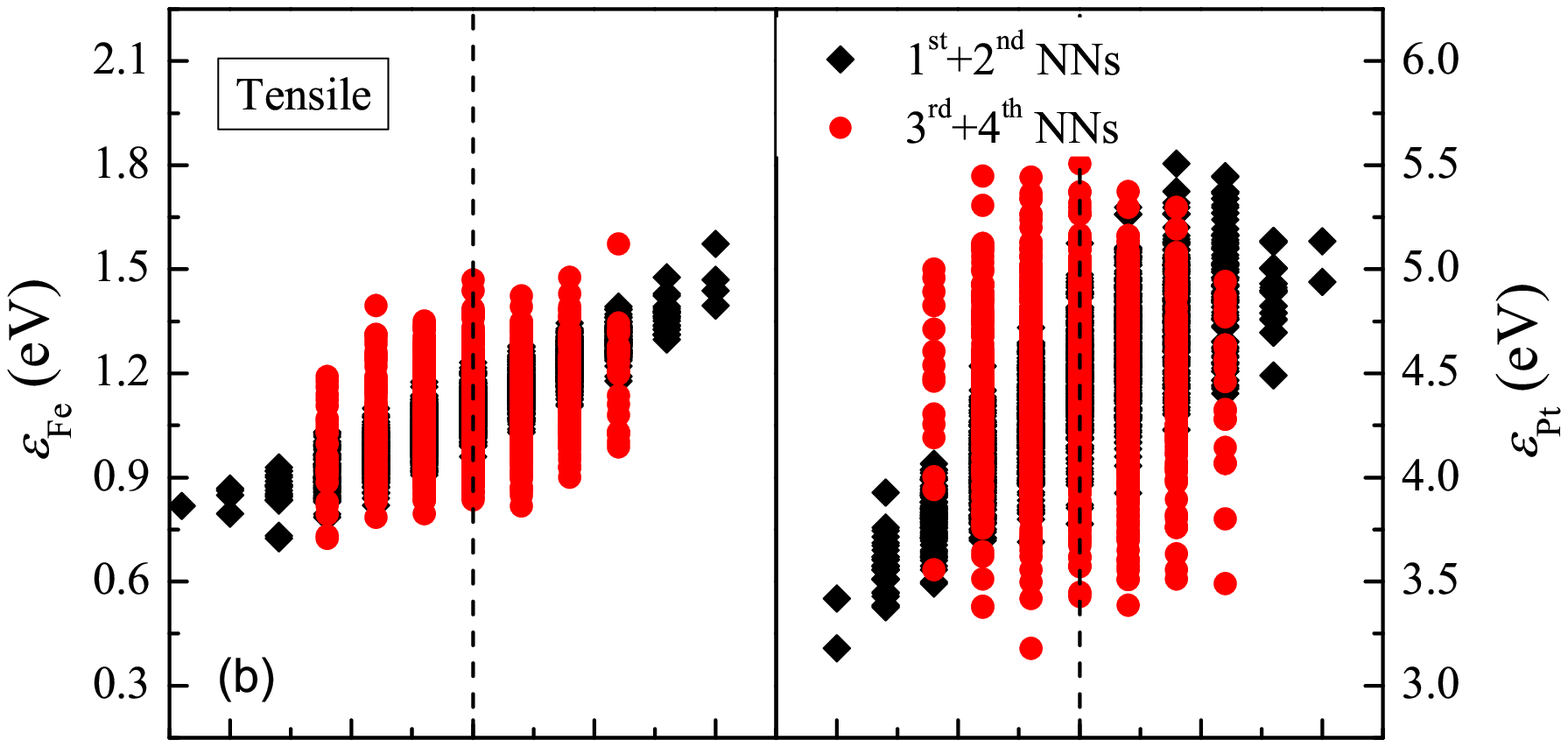}\\
\includegraphics[width=7.3cm]{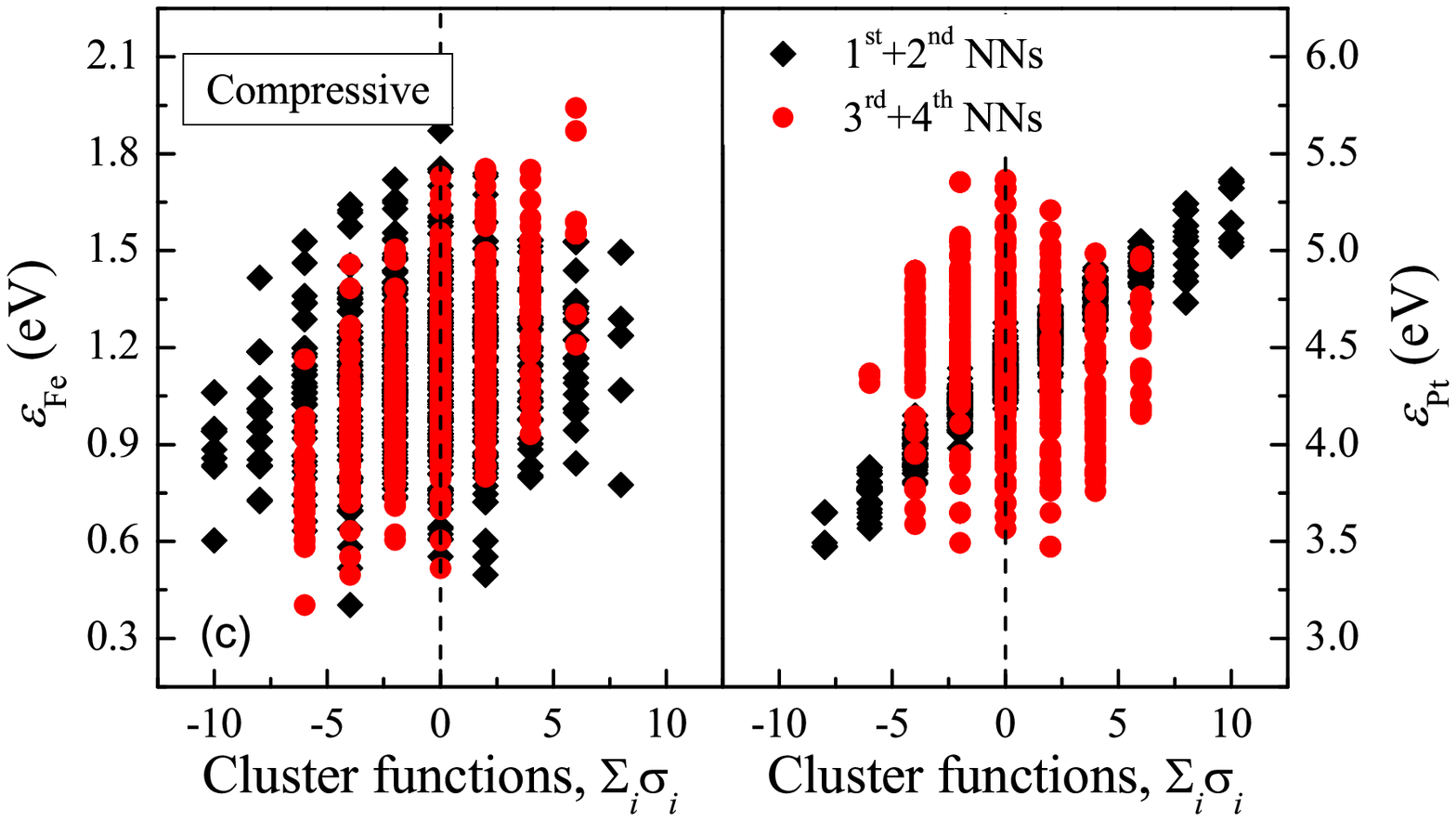}
\caption{\label{cf} (Color online) The local environment-dependent $\varepsilon_{\rm Fe}^{}$ and $\varepsilon_{\rm Pt}^{}$ (see the caption of Fig.~\ref{dos}) for Fe vacancies (left panel) and Pt vacancies (right panel), respectively, under FM state. See more in the text.}
\end{figure}

Since $\varepsilon_{\rm Fe}^{}$ and $\varepsilon_{\rm Pt}^{}$ are closely related to the local environment, their variations shown in Fig.~\ref{dos} may be further discussed by examining the atomic coordination of each site. For simplicity, we focus only on the amount of each species in the vicinity of the vacancies. This can be done by plotting $\varepsilon_{\rm Fe}^{}$ and $\varepsilon_{\rm Pt}^{}$ against the single-point cluster functions ($\sum_i\sigma_i$) of the atomic shells. A positive $\sum_i\sigma_i$ indicates an Fe-rich environment while the negative is Pt-rich. Due to the difference in symmetry, two and four atomic shells around the vacancies should be considered for the unstrained and strained alloys, respectively. However, for consistency, the $\sum_i\sigma_i$ of the strained ones is the sum over the first (third) and second (fourth) NNs, and thus we still refer to them as the first and second atomic shells for convenience, as shown in Fig.~\ref{cf}.


For each alloy, $\varepsilon_{\rm Fe}^{}$ increases when the coordination of the Fe vacancies transits from Pt-rich to Fe-rich. A similar relation is also found for $\varepsilon_{\rm Pt}^{}$ with the first atomic shell coordination. Since there is no clear dependence of $\varepsilon_{\rm Pt}^{}$ on $\sum_i\sigma_i$ in the second atomic shell, the variation of $\varepsilon_{\rm Pt}^{}$ is determined by the interaction of Pt atoms with their first NNs. Apparently, the vacancies prefer to be generated in Pt-rich local environments. This agrees well with the lower $E_{\rm{v}}^{\rm{f}}$ of elemental Pt compared to Fe (see Table \ref{vfe_L10}). However, as temperature increases, the vacancies with a locally higher amount of Fe (larger $\varepsilon_{\rm Fe}^{}$ and $\varepsilon_{\rm Pt}^{}$) can also be thermally generated, which is responsible for the increase in $\overline{E}_{\rm{v}}^{\rm{f}}$ as shown in Fig.~\ref{evac}. These discussions are independent of the magnetic state and thus the DLM cases are not shown here.


When the alloy is subject to a strain, either tensile or compressive, $\varepsilon_{\rm Fe}^{}$ decreases for the Fe vacancies with Pt-rich local environments while increases for those with Fe-rich local environments (see also Fig.~\ref{dos}). In contrast, the strains cause decrease in $\varepsilon_{\rm Pt}^{}$ for all the Pt vacancies. These results may be explained by the different responses of bond strength to the strains between different central atoms (to be vacated) and their NNs. According to the Bain's transformation path of elemental Fe (FM) \cite{Friak2001}, the Fe-Fe bond in fcc-Fe is elastically unstable. Hence, the Fe-Pt and Pt-Pt bonds should be elastically stable to keep the lattice of $A1$-FePt at least metastable at room temperature. The result is that, for all Pt-centered local environments, the strains will reduce the overall bond strength between the central Pt and its NNs and thus cause decrease in $\varepsilon_{\rm Pt}^{}$. However, for the Fe-centered local environments, the overall bond strength between the central Fe and its NNs decreases if the NNs are Pt-rich ($\varepsilon_{\rm Fe}^{}$ decreases) and increase if the NNs are Fe-rich ($\varepsilon_{\rm Fe}^{}$ increases) when applying strains. Hence, the two strain effects on $\varepsilon_{\rm Fe}^{}$ compete with each other. A reached balance explains why $\Delta\varepsilon_{\rm Fe}^{}$ increases due to the strains but the $\varepsilon_{\rm Fe}^{}$ with respect to $g_{\rm Fe}^{\rm max}$ is basically unchanged.


Since the Fe atoms in $A1$-FePt have strongly spin-polarized ground state, the magnetic state mainly influences the atomic bonds involving Fe. The Fe-Pt and Fe-Fe bonds become less stable (thus the bond strength decreases) from FM to DLM state. This is the reason why the magnetic excitation significantly lowers $\varepsilon_{\rm Fe}^{}$ but has less effect on $\varepsilon_{\rm Pt}^{}$. In addition, the Fe-Fe bonds are more sensitive to magnetic state than the Fe-Pt bonds. Therefore, from FM to DLM, more Fe NNs around Fe vacancy corresponds to a greater decrease in $\varepsilon_{\rm Fe}^{}$, which leads to the narrowing of $\Delta\varepsilon_{\rm Fe}^{}$ of $g_{\rm Fe}^{}$ (see Fig.~\ref{dos}).

In summary, we have proposed an efficient but also effective method to calculate the temperature-dependent mean vacancy formation energy (VFE) in solid solutions, based on which the effects of (001) in-plane strains and magnetism on the mean VFE of $A1$-FePt are theoretically studied. The calculated mean VFE increases with temperature but significantly decreases due to strains and can be further reduced by magnetic excitation. The vacancies prefer to be generated within Pt-rich local environments, in which the overall bond strength between the central atom and its nearest neighbors can be weakened by the strains and magnetic excitation and thus results in decrease in the mean VFE.

\section*{Acknowledgements}
The authors acknowledge the financial support from the Ningbo Scientific and Technological Project under Grant No.~2011B82004 and NSFC under Grant Nos.~51401227 and 51422106. QMH thanks the financial support from MoST of China under Grant No. 2014CB644001. JPL thanks US/DoD/ARO support under grant W911NF-11-1-0507. HBL thanks Prof. Andrei Ruban from KTH, Sweden, for helpful discussion.


\begin{thebibliography}{99}
\bibitem{Korzhavyi1999} Korzhavyi, P. A., Abrikosov, I. A., Johansson, B., Ruban, A. \& Skriver, H. L. First-principles calculations of the vacancy formation energy in transition and noble metals. \emph{Phys. Rev. B} \textbf{59}, 11693(1999).
\bibitem{Soven1967} Soven, P. Coherent-potential model of substitutional disordered alloys. \emph{Phys. Rev.} \textbf{156}, 809-813(1967).
\bibitem{Gyorffy1972} Gyorffy, B. L. Coherent-potential approximation for a nonoverlapping-Muffin-tin-potential model of random substitutional alloys. \emph{Phys. Rev. B} \textbf{5}, 2382-2384(1972).
\bibitem{Delczeg2012} Delczeg, L., Johansson, B. \& Vitos, L. Ab initio description of monovacancies in paramagnetic austenitic Fe-Cr-Ni alloys. \emph{Phys. Rev. B} \textbf{85}, 174101(2012).
\bibitem{Zhang2015} Zhang, X. \& Sluiter, M. H. F. Ab initio prediction of vacancy properties in concentrated alloys: The case of fcc Cu-Ni. \emph{Phys. Rev. B} \textbf{91}, 174107(2015).
\bibitem{Sanchez1984} Sanchez, J. M., Ducastelle, F. \& Gratias, D. Generalized cluster description of multicomponent systems. \emph{Physica A} \textbf{128}, 334-350(1984).
\bibitem{Wolverton1994} Wolverton, C. \& de Fontaine, D. Cluster expansions of alloy energetics in ternary intermetallics. \emph{Phys. Rev. B} \textbf{49}, 8627(1994).
\bibitem{Lechermann2000} Lechermann, F. \& F\"{a}hnle, M. Ab initio statistical mechanics for alloy phase diagrams and ordering phenomena including the effect of vacancies. \emph{Phys. Rev. B} \textbf{63}, 012104(2000).
\bibitem{Van2005prl} Van der Ven, A. \& Ceder, G. First principles calculation of the interdiffusion coefficient in binary alloys. \emph{Phys. Rev. Lett.} \textbf{94}, 045901(2005).
\bibitem{Van2005} Van der Ven, A. \& Ceder, G. Vacancies in ordered and disordered binary alloys treated with the cluster expansion. Phys. Rev. B 71, 054102(2005).
\bibitem{Abrikosov1997} Abrikosov, I. A., Simak, S. I., Johansson, B., Ruban, A. V. \& Skriver, H. L. Locally self-consistent Green's function approach to the electronic structure problem. \emph{Phys. Rev. B} \textbf{56}, 9319(1997).
\bibitem{Ruban2008} Ruban, A. V. \& Abrikosov, I. A. Configurational thermodynamics of alloys from first principles: effective cluster interactions. \emph{Rep. Prog. Phys.} \textbf{71}, 046501(2008).
\bibitem{Peil2012} Peil, O. E., Ruban, A. V. \& Johansson, B. Self-consistent supercell approach to alloys with local environment effects. \emph{Phys. Rev. B} \textbf{85}, 165140(2012).
\bibitem{Berry2007} Berry, D. C. \& Barmak, K. Effect of alloy composition on the thermodynamic and kinetic parameters of the $A1$ to $L1_0$ transformation in FePt, FeNiPt, and FeCuPt films. \emph{J. Appl. Phys.} \textbf{102}, 024912(2007).
\bibitem{Wiedwald2007} Wiedwald, U. et al. Lowering of the $L1_0$ ordering temperature of FePt nanoparticles by He+ ion irradiation. \emph{Appl. Phys. Lett.} \textbf{90}, 062508(2007).
\bibitem{Hsiao2009} Hsiao, S. N. et al. Effect of initial stress/strain state on order-disorder transformation of FePt thin films. \emph{Appl. Phys. Lett.} \textbf{94}, 232505(2009).
\bibitem{Wang2011} Wang, B. \& Barmak, K. Re-evaluation of the impact of ternary additions of Ni and Cu on the $A1$ to $L1_0$ transformation in FePt films. \emph{J. Appl. Phys.} \textbf{109}, 123916(2011).
\bibitem{Li2011b} Li, Y., Feng, T., Chen, Q., Huang, A. \& Chen, Z. Promotion of $L1_0$ ordering in Pt-rich nonepitaxially grown FePt films. \emph{Mater. Lett.} \textbf{65}, 2589-2591(2011).
\bibitem{Wang2013} Wang, H. et al. One-step synthesis of high-coercivity $L1_0$-FePtAg nanoparticles: Effects of Ag on the morphology and chemical ordering of FePt nanoparticles. \emph{Chem. Mater.} \textbf{25}, 2450-2454(2013).
\bibitem{Luo2014} Luo, H. B. et al. Effect of stoichiometry on the magnetocrystalline anisotropy of Fe-Pt and Co-Pt from first-principles calculation. \emph{J. Phys.: Condens. Matter} \textbf{26}, 386002(2014).
\bibitem{Luo2015} Luo, H. B. et al. On the magnetic structure of (Fe$_{1-x}$Mn$_x$)Pt: A first-principles study. \emph{J. Magn. Magn. Mater.} \textbf{378}, 138-142(2015).
\bibitem{Chepulskii2005} Chepulskii, R. V. \& Butler, W. H. Temperature and particle-size dependence of the equilibrium order parameter of FePt alloys. \emph{Phys. Rev. B} \textbf{72}, 134205(2005).
\bibitem{Elkins2005} Elkins, K. et al. Monodisperse face-centred tetragonal FePt nanoparticles with giant coercivity. \emph{J. Phys. D: Appl. Phys.} \textbf{38}, 2306(2005).
\bibitem{Belak2015} Belak, A. A. \& Van der Ven, A. Effect of disorder on the dilute equilibrium vacancy concentrations of multicomponent crystalline solids. \emph{Phys. Rev. B} \textbf{91}, 224109(2015).
\bibitem{Greiner1995} Greiner, W., Neise, L. \& St\"{o}cker, H. \emph{Thermodynamics and statistical mechanics} (Springer-Verlag, 1995).
\bibitem{Vitos2001} Vitos, L. Total-energy method based on the exact muffin-tin orbitals theory. \emph{Phys. Rev. B} \textbf{64}, 014107(2001).
\bibitem{Vitos2007} Vitos, L. \emph{Computational Quantum Mechanics for Materials Engineers} (Springer-Verlag, 2007).
\bibitem{Perdew2008} Perdew, J. P. et al. Restoring the density-gradient expansion for exchange in solids and surfaces. \emph{Phys. Rev. Lett.} \textbf{100}, 136406(2008).
\bibitem{Glensk2014} Glensk, A., Grabowski, B., Hickel, T. \& Neugebauer, J. Breakdown of the Arrhenius law in describing vacancy formation energies: The importance of local anharmonicity revealed by \textit{ab initio} thermodynamics. \emph{Phys. Rev. X} \textbf{4}, 011018(2014).
\bibitem{Kresse1996} Kresse, G. \& Furthm\"{u}ller, J. Efficient iterative schemes for \textit{ab initio} total-energy calculations using a plane-wave basis set. \emph{Phys. Rev. B} \textbf{54}, 11169-11186(1996).
\bibitem{Korhonen1995} Korhonen, T., Puska, M. J. \& Nieminen, R. M. Vacancy-formation energies for fcc and bcc transition metals. \emph{Phys. Rev. B} \textbf{51}, 9526-9532(1995).
\bibitem{Friak2001} Fri\'{a}k, M., \v{S}ob, M. \& Vitek, V. \textit{Ab initio} calculation of phase boundaries in iron along the bcc-fcc transformation path and magnetism of iron overlayers. \emph{Phys. Rev. B} \textbf{63}, 052405(2001).
\end{thebibliography}
\end{document}